# Imaging and optical properties of single core-shell GaAs-AlGaAs nanowires

Thang B. Hoang, L. V. Titova, H. E. Jackson and L.M. Smith
Department of Physics
University of Cincinnati
Cincinnati, OH

J.M. Yarrison-Rice
Physics Department
Miami University
Oxford, OH

Y. Kim, H. J. Joyce and C. Jagadish
Department of Electronic Materials Engineeering
Research School of Physical Sciences and Engineering
Australian National University, Australia

*Abstract* - We study the optical properties of a single core-shell GaAs-AlGaAs nanowire (grown by VLS method) using the technique of micro photoluminescence and spatially-resolved photoluminescence imaging.  We observe large linear polarization anisotropy in emission and excitation of nanowires

## I. Introduction

As part of the rapid development of nanotechnology, nanowires (NW) have become active components in several nano-devices [1-3]. Among the various types of NWs, III-V semiconductor (InP, GaInP, GaAs) NWs are promising candidates for development of a number of new nanoscale optical devices such as nanowire-based photodetectors, single nanowire lasers etc.,[4-5] Recently, significant advances in growth techniques have enabled the fabrication of core-shell GaAs-AlGaAs and GaAs-InGaP NWs which have much higher optical efficiency than bare GaAs NWs since nonradiative surface recombination is suppressed.  However, to date, understanding of the optical properties of core-shell GaAs-AlGas NWs is limited, especially for *single* NWs. While single nanowire spectroscopy presents significant experimental challenges, such information is essential for understanding the optical properties of these structures, since the physical size, shape and composition of the NWs may vary significantly from wire to wire and have a strong effect on their optical properties.

We study the optical properties of a single core-shell GaAs-AlGaAs nanowire grown by the vapor-liquid-solid (VLS) method [6-7].  Using PL imaging of a single nanowire in combination with polarization analysis, photoluminescence excitation and spatially and temporally resolved spectra, we examine the low temperature (10 K) optical and electronic properties of single GaAs/AlGaAs core-shell nanowires.

## II. EXPERIMENTAL DETAILS

The GaAs/AlGaAs NW sample was fabricated by the VLS method. Undoped GaAs (111)B substrates were functionalized by dipping in 0.1% poly-L-lysine (PLL) solution for 1 min. After rinsing and drying, 30 nm diameter Au nanoparticles were dispersed on the substrate surface.  GaAs nanowires were grown by horizontal flow MOCVD. Prior to nanowire formation, the substrate was annealed in-situ at 600 °C under AsH$_3$ ambient for 10 min to desorb surface contaminants and form the eutectic alloy between Au nanoparticle and Ga from the substrate. After cooling down to the growth temperature at 450 °C, Ga source gases are switched on to initiate nanowire growth. After the growth of GaAs core NW, the growth temperature was increased to 650 °C for AlGaAs shell growth. Fig. 1a shows a 45° tilted field emission scanning electron microscope (FESEM) image of the vertical NWs grown on a (111)B GaAs substrate. The NWs have pronounced tapered shape with an average diameter of ~80 nm and average length of 6-8 μm. The tapering is due to the incorporation of reaction species from the (111) surface to the side wall [8]. We therefore estimate that the GaAs core is approximately 40 nm in diameter and the AlGaAs shell is 20 nm thick. Since the exciton Bohr radius in GaAs is 12 nm we expect only very weak quantum confinement effects in these nanowires.

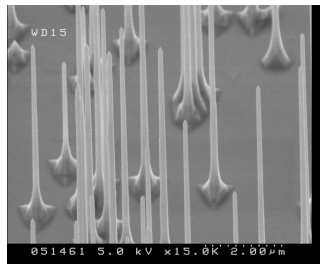

Figure 1a.  A FESEM image shows several nanowires on a GaAs substrate

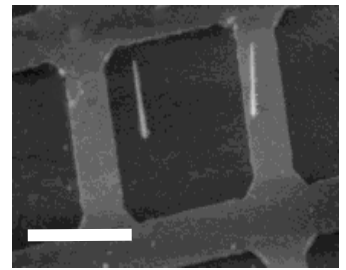

Figure 1b.  An AFM image shows two nanowires on a Si substrate with Al alignment marks. The scale bar is 10μm

In order to investigate a single NW, nanowires were removed from the growth substrate into solution and deposited onto a silicon substrate. The silicon substrate was marked with a square lattice of alignment marks for ease of identifying and keeping track of individual NWs. Fig. 1b shows an AFM image of two nanowires. The NWs were placed in a variable temperature continuous flow He optical cryostat. A 50×/0.5NA long working length microscope objective was used to project a 350× magnified image of single GaAs-AlGaAs wires onto the entrance slit of the spectrometer. Single NW PL was excited by 10mW of either 514.5nm Ar+ laser (for zero$^{th}$ order and polarization measurements) or 775 nm line of Ti:Sapphire laser for 2D CCD PL imaging; both lasers were defocused to a 10 μm spot size. The collected PL was dispersed by a 1-m focal length SPEX spectrometer with a 600 l/mm grating used in second order, and detected by a 2000×800 pixel liquid nitrogen-cooled CCD detector. All the measurements were conducted at 10K. The spatial resolution of our system is ~1.5 μm.

## II. RESULTS AND DISCUSSION

We studied a number of excitation, spatially and temporally resolved spectroscopies on single wires. Here we describe only the spatially- and polarization-resolved PL spectra from a single nanowire at 10 K. Fig. 2 shows a PL image of the NW at zero$^{th}$ order of the spectrometer. The large circle represents the defocused laser beam which is covering the entire NW. The NW is oriented along the entrance slit of the spectrometer (Y axis). From the image, it is clear that the wire emits PL more intensely at one end, possibly reflecting the wider taper (and thus larger amount of GaAs material) of the NW at one end. Since our optical resolution is only 1.5 μm, we are unable to resolve the taper of the nanowire.

Fig. 3a shows a 2D CCD image (spatial position vs. emission energy) of the nanowire. As the wire was oriented along the entrance slit of the spectrometer, the vertical axis of the image shows the spatial position along the wire, while the horizontal axis shows the emission energy. Figure 3b shows a spatial profile along the wire which was taken at 1.51 eV. Although the wire is about 6 μm long, the most intense PL emission originates from a small (~1.3 μm) portion of the NW. Figure 3c shows spectra which were taken at several positions along the wire. (The spectra have been normalized for clarity). PL spectra contain a single rather broad peak around 1.51 eV (FWHM ~25meV) which corresponds to the emission energy of free excitons in GaAs bulk epilayers. We have found the optical properties of the NW to be strongly linearly polarized. Fig. 4a) shows the PL emission intensity as a function of the linear polarization angle. For this measurement, the laser was circularly polarized. The emitted PL is clearly polarized (around 0°) in the direction along the NW axis. In Fig. 4b, the emission was analyzed for circular polarization, while the intensity of the PL emission is plotted as a function of the linear polarization direction of the laser. For maximum PL intensity, the laser polarization has a preferential direction for polarization along the nanowire axis. In these figures the solid circles (squares) are normalized data points and the solid curves are fits to $\cos^2\theta$.

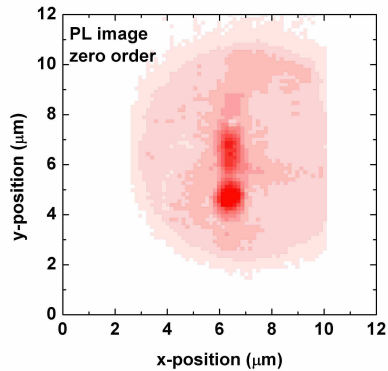

Figure 2. PL image of a nanowire at zero$^{th}$ order. The large circle represents the defocused laser beam

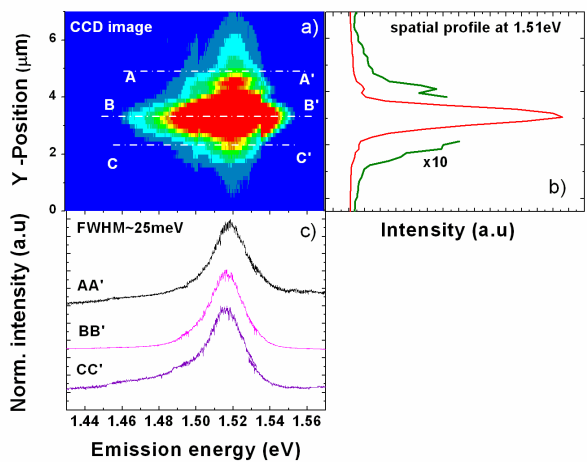

Figure 3. a) CCD image of nanowire shown in Fig..2, b) spatial profile along the wire at 1.51 eV, and c) Spectra taken at several positions along the wire

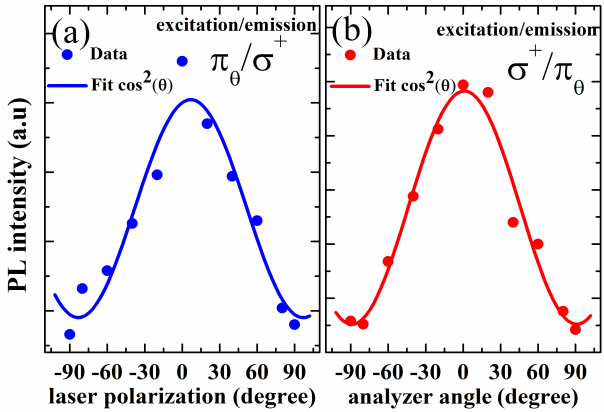

Figure 4. a) PL intensity as a function of linearly polarized excitation angles: absorption is strongly polarized along the wire, b) PL intensity as a function of linearly polarized detection angles: emission is strongly polarized along the wire

Since the quantization effects are irrelevant in the studies of this NW due to their large size compared to exciton Bohr radius, the exciton wavefunctions within the wire should be spherically symmetric and not contribute to the polarization. However, the linear anisotropy of absorption and emission may be caused by the suppression of the transverse electric field inside the NW due to the dielectric contrast between the NW and surroundings [9-10]

## CONCLUSIONS

In summary, we have studied the optical properties of single core-shell GaAs-AlGaAs nanowires. We have presented the results of the PL spectroscopy and spatially-resolved PL imaging study of one of the NWs. Zero$^{th}$ order PL imaging reveals the NW emission is spatially non-uniform, with one spot close to the wide end of the NW emitting the strongest.

We observe a single broad peak in PL centered at 1.51 eV. We found no evidence of the localized states that might be expected from defects or morphological irregularities that capture the excitons from the bulk of the wire. Also, we found that both NW emission and absorption exhibit strong linear polarization anisotropy with the preferential direction along the nanowire axis.

## ACKNOWLEDGEMENT

We acknowledge the support of the NSF through grants 0071797, and 0216374, and the Australian Research Council.

## REFERENCES

[1] Y. Gu, E.-S. Kwak, J. L. Lensch, J. E. Allen, T. W. Odom, and L. J. Lauhon., "Near-field scanning photocurrent microscopy of a nanowire photodetector," Appl. Phys. Lett., vol. 87, 043111, pp. 1-3, July 2005.
[2] Carl J. Barrelet, Andrew B. Greytak, and Charles M. Lieber., "Nanowire photonic circuit elements", Nano Lett, vol. 4, pp. 1981-1985, September 2004.
[3] Ritesh Agarwal, Carl J. Barrelet, and Charles M. Lieber., "Lasing in single cadmium sulfide nanowire optical cavities", Nano Lett, vol. 5, pp. 917-920, March 2005.
[4] Xiangfeng Duan and Charles M. Lieber., "General Synthesis of Compound Semiconductor Nanowires," Adv. Mater., vol. 12, pp. 298-302, February 2000
[5] Xiangfeng Duan, Yu Huang, Yi Cui, Jianfang Wang, and Charles M. Lieber., "Indium phosphide nanowires as building blocks for nanoscale electronic and optoelectronic devices," Nature, vol. 409, pp. 66-69, January 2001.
[6] Niklas Skold, Lisa S. Karlsson, Magnus W. Larsson, Werner Seifert, Johanna Tra1gardh and Lars Samuelson., "Growth and Optical Properties of Strained GaAs-Ga$_x$In$_{1-x}$P Core-Shell Nanowires," Nano Lett., vol. 5, pp. 1943-1947, September 2005.
[7] L. Samuelson, C. Thelander, M. T. Björk, M. Borgström, K. Deppert, K. A. Dick, A. E. Hansen, T. Mårtensson, N. Panev, A. I. Persson, W. Seifert, N. Skold, M. W. Larsson and L. R. Wallenberg., "Semiconductor nanowires for 0D and 1D physics and applications," Physica E, vol. 25, pp. 313-318, July 2004.
[8] Y. Kim, H. J. Joyce, Q. Gao, H. H. Tan, C. Jagadish, M. Paladugu, J. Zou, and A. A. Suvorova, "Influence of Nanowire Density on the Shape and Optical Properties of Ternary InGaAs Nanowires," Nano Lett., vol. 6, pp. 599-604, February 2006.
[9] Jinfang Wang, Mark S. Gudiksen, Xiangfeng Duan, Yi Cui, Charles Lieber, "Highly Polarized Photoluminescence and Photodetection from Single Indium Phosphide Nanowires," Science, vol. 293, pp. 1455-1457, August 2001.
[10] H. E. Ruda and A. Shik., "Polarization-sensitive optical phenomena in semiconducting and metallic nanowires," Phys. Rev. B, vol. 72, 115308, pp. 1-11, September 2005.